\documentclass{mem}
\usepackage{natbib}\usepackage{txfonts}\usepackage{balance}
\usepackage{graphicx}
\usepackage[a4paper]{hyperref}
\idline{xx}{1}
\begin{document}
\def\teff{$T\rm_{eff }$}
\def\kms{$\mathrm {km s}^{-1}$}
\def\la{Ly$\alpha$}
\def\hb{H$\beta$}
\def\arcmin              {$^{\prime}$}
\def\arcm                {$^{\prime}$}
\def\arcsec              {$^{\prime\prime}$}
\def\arcs                {$^{\prime\prime}$}
\def\kms                 {km\thinspace s$^{-1}$}
\def\cms                 {cm$^{-2}$}
\def\etal{{\it et al. }}
\def\aa{{\rm A$\,$\&$\,$A}}            
\def\araa{{\rm ARA$\,$\&$\,$A}}                
\def\aar{{\rm A$\,$\&$\,$AR}}          
\def\aas{{\rm A$\,$\&$\,$AS}}          
\def\apj{{\rm ApJ}}                    
\def\apjs{{\rm ApJS}}                  
\def\apjl{{\rm ApJ Let}}               
\def\aj{{\rm AJ}}                      
\def\sva{{\rm SvA}}                    
\def\pasp{{\rm PASP}}                  
\def\pasj{{\rm PASJ}}                  
\def\mnras{{\rm MNRAS}}                        
\def\kmsmpc              {km\thinspace s$^{-1}$\thinspace Mpc$^{-1}$}
\def\Msol{\thinspace\hbox{$\hbox{M}_{\odot}$}}
\def\Zsol{\thinspace\hbox{$\hbox{Z}_{\odot}$}}
\def\sol{\thinspace\hbox{$_{\odot}$ }}
\def\deg{\hbox{$^\circ$}}
\def\kpc{\thinspace\hbox{kpc}}
\def\ojo{\fbox{\bf !`$\odot$j$\odot$!}}
\def\ie{{\it i.e.} }                    
\def\a4{\hsize 17.0cm \vsize 25.cm}
\newcommand{\der}[2]  { \frac{{\rm d}#1}{{\rm d}#2} }
\newcommand{\derp}[2] { \frac{\partial #1}{\partial #2} }
\newcommand{\dif}     {{\rm d}}
\newcommand{\difp}    {\partial}
\newcommand{\Hii}{\ion{H}{2}}
\newcommand{\Sub}[1]{_\mathrm{#1}}
\newcommand{\SSub}[1]{_\mathrm{\scriptscriptstyle #1}}
\newcommand{\Sup}[1]{^\mathrm{#1}}
\newcommand{\thC}{$\theta^1$~C~Ori}
\newcommand{\vdag}{(v)^\dagger}
\newcommand{\gmtemail}{gustavo@iagusp.usp.br}

\title{SMBH Spherically Symmetric Accretion Regulated by Violent Star 
Formation Feedback}

\subtitle{}

\author{S. \,Silich, F. \,Hueyotl-Zahuantitla
\and G. \,Tenorio-Tagle}

\offprints{S. Silich}
 
\institute{Instituto Nacional de Astrof\'\i sica Optica y
Electr\'onica, AP 51, 72000 Puebla, M\'exico \\
\email{silich@inaoep.mx}
}

\authorrunning{Silich et al.}

\titlerunning{SMBH accretion regulated by violent SF}

\abstract{
The mounting evidence for violent nuclear star formation in Seyfert  galaxies 
has led us to consider the hydrodynamics of the matter reinserted by massive 
stars through strong stellar winds and supernovae, under the presence of a 
central massive BH. We show that  in all  cases there is a bimodal 
solution strongly weighted by the location of the stagnation radius ($R_{st}$),
which splits the star cluster into two different zones. Matter 
reinserted within the stagnation volume is to be accreted by the BH while its 
outer counterpart would  composed a star cluster wind. The mechanical power 
of the latter, ensures that there is no accretion of the ISM into the BH and 
thus the BH accretion and its luminosity is regulated by the star formation 
feedback. The location of the stagnation  radius is a function of three 
parameters: the BH mass, the mechanical power (or mass) of the star formation 
event and the size of the star forming region. Here we present our 
self-consistent, stationary solution, discuss the accretion rates and BH 
luminosities and show that our model predicts the intrinsic link between the 
BH activity and the starburst parameters.
\keywords{accretion --- galaxies: active --- galaxies: starburst --- 
          galaxies: star clusters --- hydrodynamics}
}
\maketitle{}

\section{Introduction}

The imprints of massive starbursts around the central supermassive black 
hole (BH) have been revealed in a number of Seyfert galaxies. 
Levenson et al. (2001), Jim\'enez-Bail\'on et al. (2005) found that in many
cases the spectral fit to the X-ray emission from Seyfert galaxies requires 
a complementary thermal component associated with an extended starburst. 
Alexander et al. (2005) provided ultra deep Chandra observations of 20 
radio-selected submillimeter-emitting galaxies selected from the deep SCUBA 
surveys and combined this with Keck deep spectroscopy. They found strong 
evidences that a substantial fraction ($\sim 75$\%) of the selected star 
forming galaxies harbor also an active galactic nucleus (AGN). They also 
found the average 
X-ray to the far-IR luminosity ratio an order of magnitude smaller than that 
in the case of a typical quasar (QSO). This led them to conclude that 
vigorous star formation activity ($\sim 1000$ \Msol \, yr$^{-1}$) dominates 
the bolometric luminosity of their AGN-classified sources and that 
submillimeter-emitting galaxies are associated with a massive black hole (BH) 
and stellar component rapid growth stage which ends up with the emergence of 
an un-obscured quasar.

Heckman et al. (1997) and Gonz\'alez Delgado et al. (1998) found 
absorption line features associated with photospheres of O and B stars and 
their stellar winds in the ultraviolet and optical spectra of four Seyfert 2 
galaxies that presented direct evidence for the existence of nuclear 
starbursts in the AGN galaxies. 

Terlevich et al. (1990) proposed to use the Ca II triplet as the near-IR 
indicator of star formation in AGN galaxies. Schweitzer et al. (2006) 
detected the polycyclic aromatic hydrocarbon (PAH) features, associated with
soft UV radiation emerging from the star forming region, in 11 from  
26 selected quasars. They found that at least 30\% but likely much more
of the far-IR luminosity in their sample QSOs arises from star formation and
concluded that there is a strong connection between the AGN and starburst 
activity in the QSOs. These results have been confirmed in Netzer et al. 
(2007) who presented the supplementing spectral energy distributions.

Seth et al. (2007) presented evidences on the presence of super massive black 
holes in galaxies which host compact nuclear star clusters. Many of these 
galaxies present a mixed AGN-starburst optical spectra and are classified as 
composite.

Strong evidences for intense high velocity outflows in the composite 
Seyfert 2/starburst ultra luminous infrared galaxies have been presented and
thoroughly discussed in  Gonz\'alez Delgado et al. (1998) and Rupke et al. 
(2005).

Thus star formation occurs at different space and energy scales around 
the SMBH in many AGN galaxies and QSOs. The mechanical power of young 
nuclear starbursts might prevent through the cluster winds the accretion of 
interstellar matter from the bulges and disks of their host galaxies onto 
the central BHs (see, for example, McLaughlin et al. 2006). Here we show that
in such cases the BHs are fed with the matter injected by numerous stellar 
winds and SNe explosions that result from the multiple evolving sources. This 
implies that nuclear starbursts must strongly affect and even control 
the power of the central BH at the stage of vigorous star formation. 
In fact, it  may be the dominant factor to be consider in order to understand
the physics of the BH growth and the relation between the BH and star burst
luminosity in AGN galaxies and QSOs. 

\section{The model}

We assume that massive stars are homogeneously distributed inside a spherical 
volume of radius $R_{SC}$ and that the mechanical energy deposited through 
stellar winds and supernovae, $L_{SC}$, is thermalized via random collisions 
of the gaseous streams  from neighboring  sources. This results into a high 
temperature and a high thermal pressure that drives a high velocity 
($u \sim 1000$ km s$^{-1}$) outflow of the injected matter.

However in presence of a massive central black hole, a fraction of the 
deposited matter is to remain bound inside the cluster to eventually fall 
onto the central BH. The outer zones, where the star burst wind is formed,
and the central accretion zone are separated by the stagnation radius,
$R_{st}$, where the expansion velocity, $u = 0$ km s$^{-1}$. The position of 
the stagnation point is an important issue because it defines both the 
accretion rate onto the central BH, and the amount of matter that the star 
forming region returns to the host galaxy interstellar medium. This is 
defined by
the two boundary conditions (see Silich et al. 2008).
Specifically, the outer and inner sonic points (the points where the speed of 
the flow is equal to the local sound speed) must be located at the star cluster
surface and in the star cluster center, respectively. Figure 1 presents a 
schematic representation of our model.
\begin{figure*}[t!]
\resizebox{\hsize}{!}{\includegraphics[clip=true]{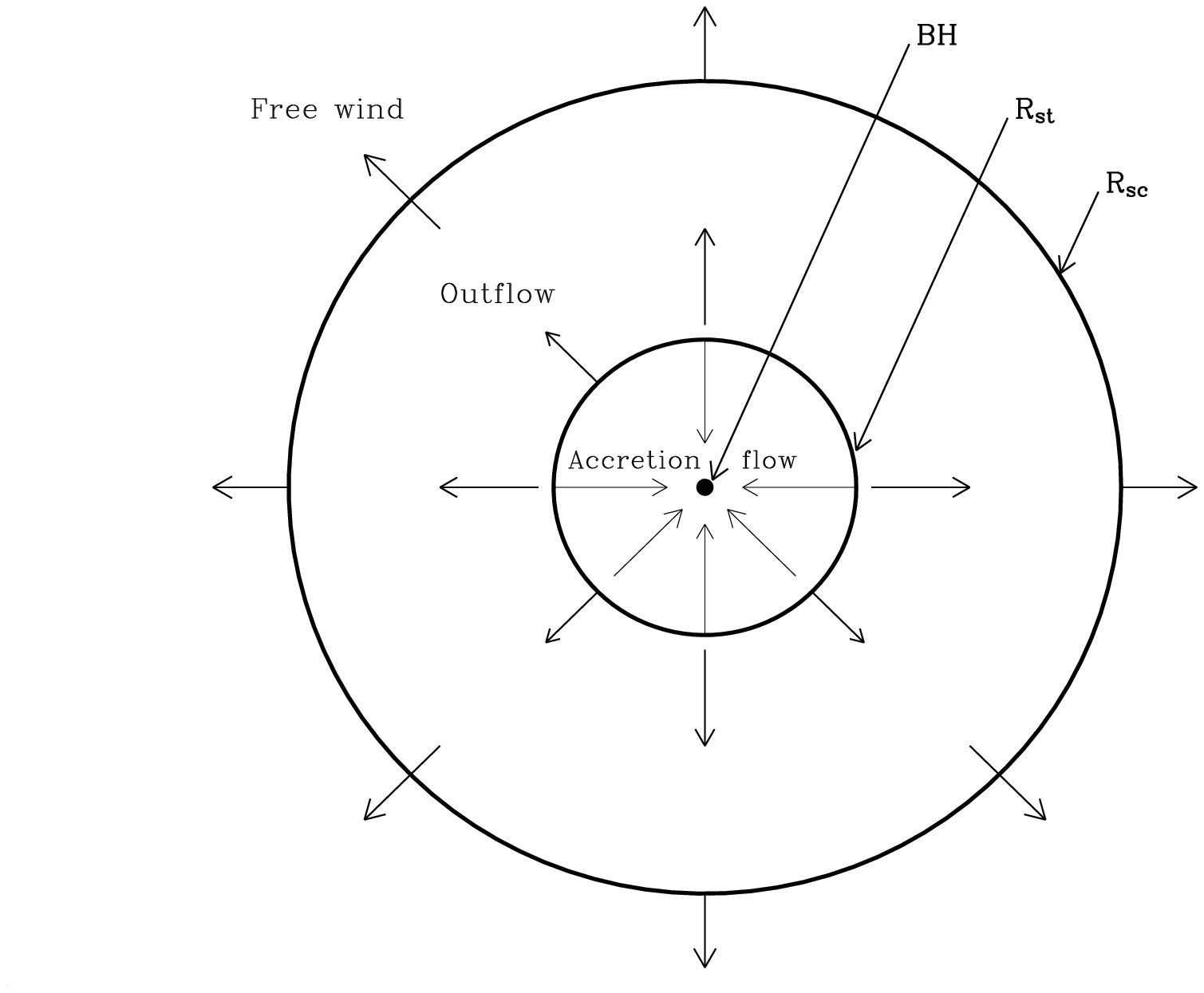}}
\caption{\footnotesize
The structure of the flow that results from the thermalization of the 
supernova ejecta and stellar winds inside a star forming region with
a massive black hole at the center. The radii of the internal 
and the external circles represent the stagnation radius $R_{st}$ and the 
star cluster radius $R_{sc}$, respectively. The arrows indicate the 
direction of the flow. The black dot at
the center marks the location of the black hole.
}
\label{Fig1}
\end{figure*}

\section{Accretion rates and BH luminosities} 

Silich et al. (2008) found that the accretion rate and the luminosity
of the BH located at the center of a young stellar cluster depend on 
the BH and star cluster mass and on the size of the 
star forming region. There is a surface in this 3D parameter space which 
separates clusters evolving in the stationary regime from those which cannot 
fulfill stationary conditions. Figure 2 presents the threshold mechanical
luminosity for stellar clusters with a $10^8$\Msol \, central BH. 
\begin{figure}[]
\resizebox{\hsize}{!}{\includegraphics[clip=true]{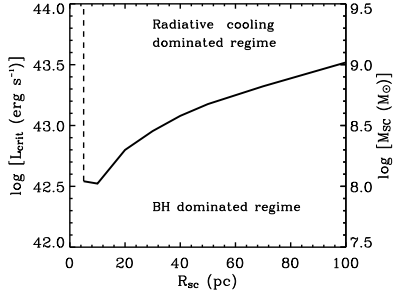}}
\caption{The threshold mechanical luminosity for a star cluster with
a $10^8$\Msol \, BH at the center. The stationary solution exists only
below the threshold line. Above the threshold line the thermalized plasma 
inside the cluster is thermally unstable and the flow has very complicated
time-dependent velocity pattern. The vertical dashed line marks the critical 
radius, $R_{crit}$. Cooling cannot compete with gravity and the stationary 
solution always exists if the star cluster radius, $R_{SC} < R_{crit}$. 
\footnotesize
}
\label{Fig2}
\end{figure}

Clusters whose mechanical luminosity (and mass) exceeds the critical value 
cannot form a stationary outflow. The thermalized plasma inside such clusters 
is thermally unstable, presents a complicated velocity pattern and is expected
to be re-processed into new generations of stars (W\"unsch et al. 2008). 

Clusters whose mechanical luminosity is smaller than the critical value, 
compose stationary accretion flow in the central zones and form stationary
outflows in the outer zones of the cluster. The accretion luminosity of a 
$10^8$\Msol \, BH located at the center of such clusters is presented 
in Figure 3. It is instructive to note that the BH luminosity can approach 
the Eddington limit only in the case of very compact stellar clusters whose 
radii are smaller than the critical value, $R_{crit}$, marked in  
Figure 2 by a vertical line.

\begin{figure}[]
\resizebox{\hsize}{!}{\includegraphics[clip=true]{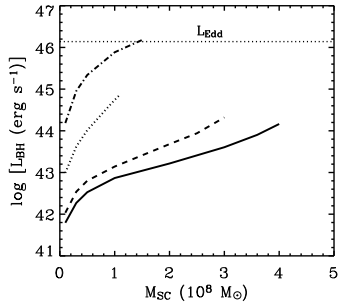}}
\caption{The BH accretion luminosity. The accretion luminosity of a 
$10^8$\Msol \, BH embedded into progressively more compact clusters.
Solid, dashes, dotted and dash-dotted lines present the results of the 
calculations for star clusters whose radii are $R_{SC} = 40$~pc, $R_{SC} = 
30$~pc, $R_{SC} = 10$~pc and $R_{SC} = 3$~pc, respectively. The horizontal 
dotted line displays the Eddington limit. The last point on every line 
displays the BH accretion luminosity when the star cluster reaches the 
threshold line. 
In the case of very compact ($R_{SC} = 3$~pc) star cluster we stopped our 
calculations when the BH luminosity reaches the Eddington value.
\footnotesize
}
\label{Fig3}
\end{figure}

\section{Conclusions}

Here we discuss the link between the supermassive BH activity and star 
formation in the central zones of AGN galaxies and QSOs. We briefly formulate 
the input 
physics and major results from our (Silich et al. 2008) self-consistent, 
stationary, spherically symmetric solution for gaseous flows which are formed 
inside starburst regions with a central supermassive BH.  

We show that the thermalization of the kinetic energy released by
massive stars inside young stellar clusters results into a bimodal solution
with an accretion of the injected matter onto a central BH in
the inner zones of the cluster, and a fast outflow of the returned matter
from the outer zones of the star forming region.

Our model predicts the intrinsic link between the BH activity and parameters 
of the starburst region. Specifically, we show that the BH luminosity,
$L_{BH}$,  depends 
on the mass and size of the stellar cluster hosting a BH. It grows with 
mass of the cluster and is larger in the case of more compact clusters.
In the case of extended starbursts, the BH luminosities fall well below 
the Eddington limit. However, $L_{BH}$ approaches the Eddington value in
the case of very compact star forming regions. 

Recent results dealing with far-IR emission of QSO and spectral properties 
of SCUBA galaxies suggest that a continuous mode of star formation is relevant
in order to understand the intrinsic links between AGNs and star formation 
activity. This will be a subject for our forthcoming communication.
 
\begin{acknowledgements}
SS thanks H. Netzer for his valuable comments and useful discussion on
our study. We thanks the organizers of the meeting for their efforts
allowing us to present and discuss our results in advance of publication.
This project has been supported by CONACYT - M\'exico research grants 
47534-F and 60333.
\end{acknowledgements}

\bibliographystyle{aa}

\end{document}